\documentclass[
 reprint,
 amsmath,amssymb,
 aps, physrev,
floatfix,
]{revtex4-2}
\usepackage{amsmath,amsfonts,amssymb,amsbsy}
\usepackage{mathrsfs,latexsym}
\usepackage{graphicx}
\usepackage{xcolor}
\usepackage{dcolumn}
\usepackage{tikz}
\usepackage[bookmarks,
bookmarksopen = true,
bookmarksnumbered = true,
linktocpage,
colorlinks = true,
linkcolor = blue,
urlcolor  = blue,
citecolor = blue,
anchorcolor = green,
hyperindex = true,
hyperfigures]{hyperref}
\usepackage{array}
\usepackage{booktabs}
\usepackage{multirow}
\usepackage{placeins}
\usepackage{blindtext}
\usepackage{appendix}
\usepackage{fontawesome} 
\usepackage{bm}
\usepackage{times}
\usepackage{wrapfig}
\usepackage{physics}
\usepackage[utf8]{inputenc} 
\usepackage[T1]{fontenc} 
\usepackage[english]{babel}
\preprint{WUB/26-00}
\begin{document}

\begin{flushright}
\textbf{WUB/26-00}
\end{flushright}

\title{\textbf{Charmonium - Glueball spectroscopy with improved hadron creation operators } 
}%

\author{Juan Andrés Urrea-Niño}
\affiliation{School of Mathematics, Trinity College Dublin, Dublin 2, Ireland}
\affiliation{Hamilton Mathematics Institute, Trinity College Dublin,
Dublin 2, Dublin, Ireland}
\affiliation{Department of Physics, Bergische Universit\"at Wuppertal, Gau{\ss}str. 20, 42119 Wuppertal, Germany}
\author{Francesco~Knechtli} 
\author{Tomasz~Korzec} 
\affiliation{Department of Physics, Bergische Universit\"at Wuppertal, Gau{\ss}str. 20, 42119 Wuppertal, Germany}

\author{Michael Peardon}
\affiliation{School of Mathematics, Trinity College Dublin, Dublin 2, Ireland}
\affiliation{Hamilton Mathematics Institute, Trinity College Dublin,
Dublin 2, Dublin, Ireland}

\date{\today}

\begin{abstract}
Construction of creation operators which can properly sample the underlying energy eigenstates remains a fundamental first step in lattice QCD spectroscopy calculations, particularly when the spectrum includes states with different composition such as mesons, glueballs, multi-particle states, etc. We tackle this issue in the study of the scalar glueball and charmonium mixing, where we use improved operators for both types of states to resolve the low-lying spectrum and identify the dominant composition of each state in a mass regime where the glueball is stable. We include derivative-based meson operators combined with distillation profiles, as well as glueball operators built from the chromo-magnetic field and its derivatives which retain angular momentum information from their continuum counterparts. We comment on the advantages of these operators, particularly on the construction and implementation of the glueball ones, thanks to which we identify the lightest iso-scalar state as glueball-dominated $0^{++}$.
\end{abstract}

\maketitle


\section{\label{sec:level1}Introduction}
Whether glueballs, hadrons composed mainly of gluons \cite{Fritzsch:1972jv, Fritzsch:1975tx, Jaffe:1985qp} exist, remains one of the most significant open questions both experimentally and theoretically for the standard model. These exotic bound-states can share quantum numbers with conventional hadrons \cite{Crede-2009, Ochs:2013gi, ParticleDataGroup:2024cfk}, like flavor-singlet mesons, and most studies predict glueballs to be heavier than quark-model states, e.g. the lightest scalar glueball is expected around $1700$ MeV \cite{Peardon-1999} while the lightest iso-scalar scalar meson is the $f_0(500)$ below $1$ GeV \cite{ParticleDataGroup:2024cfk}. While the common quantum numbers lead to possible mixing between glueballs and conventional meson states, their predicted heavy masses open a wide array of possible decay channels \cite{Cheng2006, Cheng2015, Janowski-2014, Frere-2015}. A recent identification of a possible pseudo-scalar glueball candidate added further interest to glueball studies \cite{Ablikim-2024}. The lattice quantum chromodynamics (QCD) community has been a key participant on the theoretical side via calculations from first principles. Landmark studies of glueballs in a simplified theory without quarks \cite{Morningstar1999} were an initial approach to study them and at present most lattice QCD calculations include the dynamics of quarks and gluons together in the simulations, which is a fundamental ingredient to understand the mixing between the different states and their decays. Two recent reviews emphasizing lattice QCD calculations are in \cite{Morningstar-2025, Vadacchino-2023}.
\par
There are still other obstacles to overcome. Lattice QCD calculations rely on Monte Carlo averages. Several of these quantities suffer from a signal-to-noise problem which is more severe in the study of glueballs \cite{Parisi-1984, Lepage:1989hd}. Energies of different states of interest are extracted from the exponential decay behavior of euclidian two-point correlation functions at large enough temporal separations, however these become too statistically noisy before this asymptotic regime is reached. Significant work is already being invested into tackling the signal-to-noise problem via the multi-level sampling technique for mesonic and gluonic observables \cite{Meyer:2002cd, Meyer:2003hy, Luscher:2001up, Ce:2016ajy, Ce:2016idq, Ce:2017ndt, Ce:2019yds, Majumdar:2003xm, Majumdar:2014cqa, Luscher:1993xx, Giusti:2017ksp, Barca2024, Barca2025}. The final goal is to use this technique in the presence of dynamical quarks, where understanding the mixing dynamics between mesons and glueballs can shed some light on the nature of these exotic states. The short-time behavior of the correlation function, where statistical uncertainty is still small and therefore we can extract useful physical information, is affected by a different problem. Here, different states with identical quantum numbers contribute to the correlation function with different exponentially decaying terms. Since the temporal separation is not sufficiently large for the lightest state with slowest decaying exponential to dominate, we often cannot reliably extract the energy of the lightest state. The coefficients of these different terms are dictated by how well the hadron creation and annihilation operators used in the correlation function sample the true energy eigenstates of the theory. Operators which mostly resemble only the state of interest and not any others will lead to a faster convergence to a single exponential decay and therefore to a reliable extraction of the corresponding energy at short time separations. While there are some prescriptions on how to choose operators that are somewhat close to satisfying this condition, it is not possible to know a priori the "perfect" operator and additional techniques, like the formulation of a generalized eigenvalue problem (GEVP) to help determine it \cite{Luscher, Blossier}, are almost always used. 
\par
In this work we continue our efforts to study the scalar glueball \cite{Urrea-Scalar} albeit in a theory where this state is expected to be the lightest one. We extract the iso-vector and iso-scalar $J^{PC}=0^{++}$ spectrum in $N_f = 2$ QCD with two degenerate clover-improved Wilson quarks, each one at half the physical charm quark mass. With an iso-vector pseudo-scalar ("pion") of mass $\approx$ 2.2 GeV, a scalar glueball around $1.7$ GeV is well below the 2-pion threshold. This makes the state stable while also ensuring a strong exponential suppression of the contributions of two-particle states to the correlation functions. We pay special attention to the operator construction for both mesons and glueballs, choosing those with continuum-like structure similar to the method put forward by the Hadron Spectrum Collaboration (HadSpec) \cite{Dudek-2008, Dudek-2010} to better sample the energy eigenstates of the theory. For the glueball-like operators we use the continuum-like construction in terms of the chromo-magnetic component $\mathcal{B}_i$ and its gauge-covariant spatial derivatives $D_i$ \cite{Jaffe:1985qp, Chen-2005}.

\section{Lattice ensemble details}
We use an ensemble, denoted Em1, with lattice size $48\times 24^3$ and two degenerate dynamical non-perturbatively $O(a)$-improved Wilson quarks \cite{Jansen-1998, Jansen-2002} with the mass set to half the physical charm quark mass, periodic boundary conditions for the gauge links and Wilson plaquette gauge action. The hopping parameter is $\kappa = 0.13270$ and the bare coupling is $g_0^2 = \frac{6}{5.3}$, resulting in a lattice spacing $a = 0.0658(10)$ fm \cite{Fritzsch-2012, Cali-2019} and flow scale \cite{Luscher-2010} $\frac{t_0}{a^2} = 1.8486(7)$. We use the technique of optimal distillation profiles which we presented in \cite{Urrea-2022} using this same ensemble and so we reuse the distillation data we calculated there, i.e. perambulators and elementals. Algorithmic details on how the distillation vectors and perambulators were calculated can be found there. As in that study, we use $N_v = 200$ distillation vectors. The Em1 ensemble has around $30,000$ gauge configurations, however we measured perambulators only on a subset of $4080$ configurations. Nonetheless we will also report on the glueball measurements performed on the full statistics, since these require no inversions of the Dirac operator and are therefore relatively cheap to evaluate.

\section{Operator construction}
The GEVP formulation \cite{Luscher, Blossier, Berg-1983} remains the most robust approach to extracting the energy spectrum based on building a temporal correlation matrix from a set of linearly independent lattice hadron creation operators sharing the same quantum numbers. In the absence of statistical noise and in the limit of infinitely large temporal separation, this approach can resolve $N$ distinct energy levels using $N$ operators as long as the energy states have non-zero overlaps with the states created by the operators. However, in realistic calculations there are additional complications. The signal-to-noise problem present in the correlation matrix directly impacts the GEVP and the uncertainties of extracted energies of states higher up the ladder of excitations usually becomes increasingly large. In turn, the signal for these high enough states is lost before a reliable plateau can be found. While the reference time of the GEVP can be systematically chosen for faster convergence of the effective masses \cite{Blossier}, this requires going to larger values of time separation, which is also challenging since the correlation matrix is not necessarily positive-definite once the statistical noise is large. Furthermore,
if the operators used for the GEVP do not have a large enough overlap with all states in the spectrum of interest then one or more could be missed by the GEVP, i.e. the extracted finite-volume spectrum lacks one or more states. To avoid this, lattice calculations attempt to saturate the spectrum by including operators with same quantum numbers but different number of (anti-)~quarks \cite{Shrimal-2025, Barca-2023, Prelovsek-2025, Yan-2025}, gauge covariant derivatives \cite{Liao-2002, Gattringer-2008,Dudek-2008, Dudek-2010, Yan-2025},  quark and gauge link smearing levels , etc.  
\subsection{Meson operators}
Here we focus on a specific strategy widely used by the Hadron Spectrum collaboration to build one-particle meson operators with fixed lattice angular momentum $R=A_1, A_2, E, T_1, T_2$ starting from operators having fixed $SO(3)$ angular momentum $J=0,1,2,3,...$\cite{Dudek-2008, Dudek-2010}. Since the cubic group is a subgroup of $SO(3)$, operators which generate an irreducible representation (irrep) labeled $J$ in $SO(3)$ generate a representation, not necessarily irreducible, of the cubic group \cite{Bunker-1998, Altmann-1994}. Some examples of this \textit{subduction} relation are $0\rightarrow A_1$, $1 \rightarrow T_1$, $2 \rightarrow E \oplus T_2$. Operators of a lattice irrep $R$ are also labeled by the continuum $J$ from which they were built and are expected to mostly overlap with that same $J$ as the continuum is approached. The overlaps in the spectral decomposition of correlations built from operators in different lattice irreps coming from a common $J$ continuum operator are related to each other and this serves to identify the dominant $J$ for each energy level across all lattice irreps \cite{Dudek-2010}. For example, with this approach we can tell which states in the ladder of excitations of the $A_1$ irrep correspond to $J=0$, $J=4$ and higher values of $J$. This is particularly useful for charmonium studies where the $1^{--}$ states $J/\psi$, $\psi(2S)$ and $\psi(3770)$ lie below the $3^{--}$ state $\psi_3(3842)$ which is then followed by the $1^{--}$ $\psi(4040)$; telling apart $J=1$ and $J=3$ is fundamental to disentangle these states in a lattice calculation. The HadSpec approach not only resolves the ladder of excitations but also disentangles the dominant $J$ and identifies \textit{hybrid} states containing explicit gluonic excitations beyond a simple $\bar{c}c$ structure \cite{Liu-2012}. 
\par
For this work we start with some of the HadSpec meson operators presented in \cite{Dudek-2008} for the $A_1^{++}$ irrep. Namely, we take $\bar{c}\Gamma c$ as our charmonium operators with $\Gamma = \mathbb{I}, \gamma_i \nabla_i, \gamma_0 \gamma_5 \gamma_i \mathbb{B}_i$, where $\nabla_i$ is the gauge-covariant spatial derivative in direction $i$ and $\mathbb{B}_i = \epsilon_{ijk} \nabla_j \nabla_k$. We note that $\mathbb{B}_i$ and $\mathcal{B}_i$ as will be defined later are two different objects. By using up to two spatial derivatives, we expect to sample different spatial structures which are relevant for radial excitations. Furthermore, using the \textit{chromo-magnetic} component $\mathbb{B}_i$ we can better sample non-trivial gluonic contributions to the different excitations. These three operators are even under time reversal and therefore can be put together in a single correlation matrix without introducing additional problems in the GEVP when using periodic boundary conditions in time for the gauge links~\cite{Bailas-2018}. While these three operators have the above mentioned useful features, we need more in order to reliably extract the spectrum. Instead of including more derivative-based operators to increase our basis, we introduce meson distillation profiles in each operator as was first shown in \cite{Urrea-2022}. As in that previous study, we use seven gaussians with different widths but in this study three different $\Gamma$ operators are included. This combination of approaches yields $21$ different operators. It is worth noting that in \cite{Liu-2012} the use of up to three spatial derivatives yielded a total of $13$ operators for the $A_1^{++}$ channel using standard distillation \cite{Peardon-2009}. The use of profiles is computationally much cheaper as it does not require building increasingly complicated derivative combinations and it almost doubles the total number of operators compared to a basis involving up to three derivatives. The most complete approach would be to use all $13$ derivative-based operators together with profiles for each one, however the size of the correlation matrix would become too large. Fortunately, pruning procedures to reduce the basis size while retaining the physically relevant information are well-known~\cite{Balog-1999, Niedermayer-2001, Blossier} . For this work we are interested in the very low-lying spectrum and therefore our choice of $21$ operators is sufficient to be an efficient proof-of-principle of the combined approach for operator construction.
\subsection{Glueball operators}
Glueball operators are conventionally built from spatial Wilson loops of different shapes with different levels of link smearing~\cite{Berg-1983,Sun-2018,Sakai-2023, Barca2024}. While it is easy to build a wide basis of different loop shapes, spatial link smearing has been shown to make the resulting operator basis almost degenerate ~\cite{Sakai-2023}. This makes the correlation matrix close to singular, the GEVP becomes ill-posed and pruning become necessary. Alternative constructions are based on lattice versions of operators involving the full field-strength tensor $F_{\mu \nu}$ \cite{Sun-2018}, however this involves temporal links which are unaffected by spatial link smearing. Inspired by the subduction-based construction of meson operators previously described, we use operators built from the \textit{chromo-magnetic} component of the field-strength tensor $\mathcal{B}_i \propto \epsilon_{ijk}F_{jk}$ and its spatial gauge-covariant derivatives $D_i$ as first presented in~\cite{Chen-2005, Liu-2000, Liu-2001, Liu-2002}. The analogous 4-dimensional construction was first used to qualitatively predict the glueball spectrum based on mass-dimension arguments \cite{Jaffe:1985qp}. Following a similar procedure to the meson case, $\mathcal{B}_i$ (mass-dimension 2) and $D_i$ (mass-dimension 1) can be written in a basis with definite $J$ under $SO(3)$ and the corresponding Clebsch-Gordan coefficients used to build operators with an increasing number of $\mathcal{B}_i$ and $D_i$ in them. The simplest example for the $J^{PC}=0^{++}$ case is $1 \otimes 1 \rightarrow 0$ via the fully symmetric contraction. The resulting operator is given by
\begin{align}
    \mathcal{O}(t) &= \sum_{\vec{x}} \sum_{i}\text{Tr} \left( \mathcal{B}_i(\vec{x},t)^2  \right),
\label{eqn:ScalarOp1}
\end{align}
where we include the projection to zero spatial momentum as well as the trace to make the operator gauge-invariant. We note that this trace could be replaced by a left and right multiplication by 3D Laplacian eigenvectors in a similar spirit as in distillation, e.g. $\text{Tr}\left( \mathcal{B}(\vec{x},t)^2\right) \rightarrow \sum_{n=1}^{N_v}g_nv_n(\vec{x},t)^{\dagger} \mathcal{B}(\vec{x},t)^2v_n(\vec{x},t)$, where $g_n$ is a distillation profile. For simplicity in a proof-of-concept calculation, we use the operators with the trace, however we will also test the eigenvector approach in the future. The operator in Eq. \ref{eqn:ScalarOp1} was also discussed in \cite{Sun-2018}. Examples of operators up to mass-dimension 6 are presented in \cite{Chen-2005}, with an example of a mass-dimension 6 derivative-based operator being
\begin{align}
    \mathcal{O}(t) &= \sum_{\vec{x}} \sum_{i} \text{Tr}\left( \mathcal{B}_i(\vec{x},t) \sum_{k} D_k^2 \mathcal{B}_i(\vec{x},t)  \right).
    \label{eqn:ScalarOp2}
\end{align}
\par 
This approach has several advantages compared to the Wilson loop construction. First, the subduction retains information about the continuum $J$ as in the HadSpec approach for meson operators. Second, getting all possible operators for a fixed mass-dimension with $\mathcal{B}_i$ and $D_i$ provides a systematic way to build a basis based on the physical intuition of sampling gluonic excitations by explicitly including the field-strength tensor. Third and particularly important for the $A_1^{++}$ channels, the use of derivatives $D_i$ introduces relative minus signs and additional spatial structure which might better probe the underlying energy eigenstates. This is contrary to the case of spatial Wilson loops where we build $A_1$ operators by simply summing over all 24 cubic transformations of a fixed loop shape and there are no relative minus signs. Fourth, in this approach there is a better grasp on the linear independence between different operators. As noted in \cite{Sun-2018}, the classical small-$a$ expansion of the loop shapes used there for the $A_1^{-+}$ has the same leading term $\epsilon_{ijk} \text{Tr}\left( B_i D_j B_k \right)$ (omitting spatial and time variables as well as momentum projection). A better approach to ensure linear independence between operators is to use this mass-dimension 5 operator and then build the next possible one in odd mass-dimension, rather than use different loop shapes with equal content in the expansion. Fifth, the projection of these operators into the corresponding irreps is much simpler to implement. When using spatial Wilson loops, we have to determine which irreps, and how many times, are contained in the representation of the cubic group generated by a fixed loop shape. Only then can we sum over the 24 cubic transformations of the loop shape with specific projection coefficients~\cite{Berg-1983, Peardon-1999, Sun-2018, Sakai-2023, Barca2024, Bunker-1998}. With the approach used here, the operators are defined already in a fixed irrep and there is no need to measure different orientations of a given loop shape which then need to be further processed to build glueball operators. This is particularly relevant for the case of non-zero spatial momentum, e.g. as needed in \cite{Abbott-2026}, when the symmetry group is further broken down into one of the momentum little groups. One way to build the corresponding glueball operators is to again determine the often-reducible representations generated by a given shape orientation in the new group and perform the corresponding projection onto fixed irreps. The continuum-like operators used here already generate irreps of the cubic group and therefore the subduction into little groups can be done via tabulated relations \cite{Altmann-1994, Bunker-1998}. Finally, it is worth noting that $\mathcal{B}_i$ and $D_i$ are composed of spatial link variables, similar to spatial Wilson loops and therefore are suitable for multi-level calculations which exploit locality in time of the observables, e.g. \cite{Barca2024}.
\par
While we use the continuum-like operators presented in \cite{Chen-2005}, we implement them differently. In that work and \cite{Sun-2018}, the operators are defined on the lattice by building a linear combination of spatial Wilson loops which recovers the continuum expression up to a given power of the the lattice spacing $a$. In this work, we use the clover definition of $F_{\mu \nu}$ and set $\mathcal{B}_i = \frac{1}{2} \epsilon_{ijk} F_{jk}$ and discretize the first derivative and Laplacian acting on an a field $\phi$ with adjoint charge as seen in \cite{Lee-2002}
\begin{align}
    D_i \phi(\vec{x},t) &= \frac{1}{2} \left( U_i(\vec{x},t) \phi(\vec{x} + a\hat{i},t) U_i(\vec{x},t)^{\dagger} \right. \nonumber\\
    &- \left.  U_i(\vec{x}-a\hat{i},t)^{\dagger} \phi(\vec{x} - a\hat{i},t) U_i(\vec{x}-a\hat{i},t)\right)\\
    \sum_{k} D_k^2 \phi(\vec{x},t) &= -6\phi(\vec{x},t) \nonumber \\
    &+ \sum_{k} \left( U_k(\vec{x},t) \phi(\vec{x}+a\hat{k},t) U_k(\vec{x},t)^{\dagger} \right. \nonumber \\
    &+ \left. U_k(\vec{x}-a\hat{k},t)^{\dagger} \phi(\vec{x}-a\hat{k},t) U_k(\vec{x}-a\hat{k},t)\right).
\end{align}
Furthermore, we re-derive the operators from \cite{Chen-2005} in a different way. Instead of starting from the continuum $J$ forms and their corresponding products, we build products directly from lattice irreps. While this approach loses information from continuum contributions, the advantages previously mentioned remain. This change makes the systematic construction of operators of increasingly higher mass-dimensions more straightforward. This calculation can be done by hand and has been cross-checked using the Python package \textit{opbasis}~\cite{Husung-2025}.
\par 
Here we show our procedure and conventions for the operator construction, the latter following closely the one used in \cite{Dudek-2010}. Both $\mathcal{B}_i$ and $D_i$ generate the $T_1$ irrep in the "vector" basis, i.e. where the matrix elements correspond to the transformations of a three-dimensional vector \cite{Altmann-1994}. We build operators with definite $R^{PC}$ and label them as below, following the conventions of \cite{Dudek-2010}:
\begin{align}
    \mathcal{O}_{R,k} &= \left(  \mathcal{B}^{[1]}_{T_1} \times D^{[n_D]}_{R_D} \times \mathcal{B}^{[n_B]}_{R_B}  \right)_{R,k},
\end{align}
where $R$ is the irrep generated by the operator and $k$ is the row in the irrep. $\mathcal{B}^{[n_B]}_{R_B}$ means we couple $n_B$ $\mathcal{B}$ fields and project the result onto irrep $R_B$ and $D^{[n_D]}_{R_D}$ means we couple $n_B$ derivatives and project the result onto irrep $R_D$. In this work we use $n_D, n_B \leq 2$ and always keep the leftmost $\mathcal{B}_i$ field on its own. However, more general constructions are possible when going to sufficiently high mass-dimension. With this notation, $\left(  \mathcal{B}^{[1]}_{T_1} \times \mathcal{B}^{[1]}_{T_1}  \right)_{A_1,1}$ corresponds to Eq. \ref{eqn:ScalarOp1} and $\left(  \mathcal{B}^{[1]}_{T_1} \times D^{[2]}_{A_1} \times \mathcal{B}^{[1]}_{T_1}  \right)_{A_1,1}$ to Eq. \ref{eqn:ScalarOp2}. Following Ref.~\cite{Dudek-2010} for the $SO(3)$ case, we can write our operators in terms of the corresponding Clebsch-Gordan coefficients coupling the $\mathcal{B}_i$ and $D_i$, e.g.
\begin{align}
       \left(  \mathcal{B}^{[1]}_{T_1} \times \mathcal{B}^{[1]}_{T_1}  \right)_{R,k} &= \sum_{i,j} \braket{T_1,i;T_1,j}{R,k} \text{Tr} \left(  \mathcal{B}_i \mathcal{B}_j \right).
\end{align}
In our choice of basis the Clebsch-Gordan coefficients for $T_1 \otimes T_1 = A_1 \oplus E \oplus T_1 \oplus T_2$ can be found in \cite{Dudek-2008, Altmann-1994, Liao-2002}. For the $A_1$ we use the fully symmetric combination $\delta_{ij}$, for $T_1$ we use the anti-symmetric combination via the Levi-Civita symbol $\epsilon_{ijk}$, for $T_2$ we use $|\epsilon_{ijk}|$ and for $E$ we use the $\mathbb{Q}_{ijk}$ coefficients shown in \cite{Dudek-2008}. In the case of $\epsilon_{ijk}$, the cyclic property of the trace makes the operator vanish. The remaining $T_2$ and $E$ cases reproduce the results in \cite{Chen-2005}. For the case of three fields, for example $\mathcal{B}_i \times \mathcal{B}_i \times \mathcal{B}_i$, we have
\begin{align}
\left(  \mathcal{B}^{[1]}_{T_1} \times \mathcal{B}^{[2]}_{R_B}  \right)_{R,k} &= \sum_{i,j,k_B} \braket{T_1,i;R_B,k_B}{R,k} \nonumber\\
& \braket{T_1,i;T_1,j}{R_B,k_B} \text{Tr} \left(  \mathcal{B}_i \mathcal{B}_j \mathcal{B}_k\right).    
\end{align}
When we have more than one derivative, we follow the prescription of \cite{Dudek-2010} and couple all of them into a definite irrep before coupling this result with a $\mathcal{B}_i$, e.g.
\begin{align}
\left(  \mathcal{B}^{[1]}_{T_1} \times D^{[2]}_{R_D} \times \mathcal{B}^{[1]}_{T_1}  \right)_{R,k} &= \sum_{i,j,k_D,k_1} \braket{T_1,i;R_1,k_1}{R,k}\nonumber\\ &\braket{R_D,k_D;T_1,j}{R_1,k_1} \nonumber \\
&\braket{T_1,m;T_1,n}{R_D,k_D}\nonumber \\ 
&\text{Tr}\left(  \mathcal{B}_i D_{m} D_n \mathcal{B}_j \right).
\end{align}
Since $T_1 \otimes T_1 = A_1 \oplus E \oplus T_1 \oplus T_2$ then $R_1$ can only be one of these irreps. Furthermore, it must be chosen such that $R$ is contained in $T_1 \otimes R_1$. To extract the $T_1$ component from $E \otimes T_1$ we use the $\mathbb{R}_{ijk}$ coefficients presented in \cite{Dudek-2008} and for $T_1$ component in $T_1 \otimes T_2$ the ones in \cite{Liao-2002}. As the number of derivatives and $\mathcal{B}$ fields increase, so do the possible intermediate couplings to obtain a given $R$.
\par
\begin{figure}
    \centering
    \includegraphics[width=0.95\linewidth]{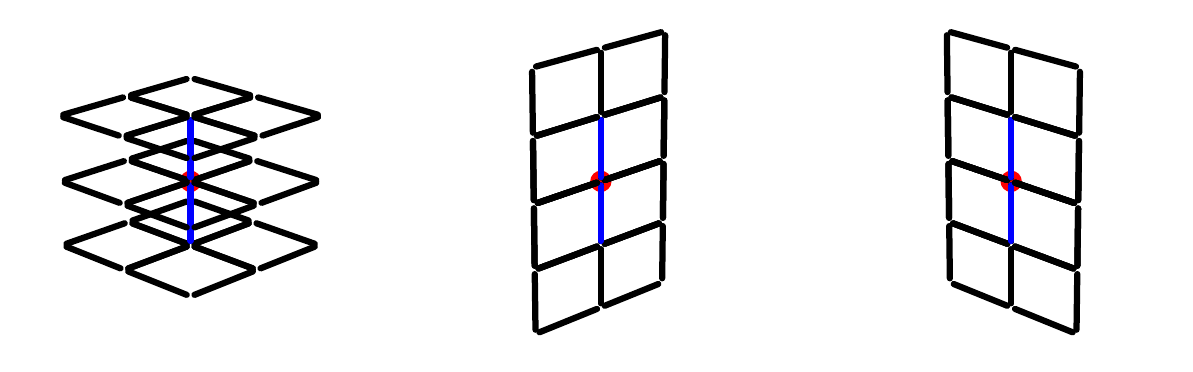}
    \caption{Link variables involved in the calculation of $D^2_k B_i(\vec{x},t)$ for a fixed spatial derivative direction $\hat{k}$ and all components $i$. The red dot marks the spatial position $\vec{x}$ and the blue links indicate the links used for the derivative.}
    \label{fig:DiB}
\end{figure}

We include some final words about our choice of implementation for these operators. First, the choice of the clover discretization of $F_{ij}$ and the symmetric finite difference for $D_i$ makes parallelization along the spatial dimensions much simpler. If we were to discretize our operators via linear combinations of different loop shapes, we would have to account for possibly complicated ones requiring more than nearest-neighbor communication. Since we only require a lattice definition of $\mathcal{B}_i$ and $D_i$ regardless of the order of the discretization error, this choice is very convenient. Second, non-trivial spatial structure in the operators is achieved by including derivatives. When working with Wilson loops we would need to implement increasingly more complicated loop shapes, which relates to the issue of spatial parallelization, while in our case we simply need to include more derivatives in the definition of the operator relying only on nearest-neighbor communication. An illustrative example is the operator given in Eq. \ref{eqn:ScalarOp2}. Each derivative term in the sum over directions of $\mathcal{B}_i$ has the Laplacian, which linearly and gauge-covariantly combines the field at spatial positions $\vec{x}$, $\vec{x}+a\hat{k}$ and $\vec{x}-a\hat{k}$. Fig. \ref{fig:DiB} shows all the link variables involved in a single term of the derivative expression for all components of $\mathcal{B}_i$. The red dot marks the position $\vec{x}$ and the blue lines indicate the links required for the derivative. Thanks to the sum over the three spatial directions, the operator uses several links in a $4a\times 4a$ cube. This large extent, together with the relative minus signs in the derivatives, gives the operator a non-trivial spatial structure. 

\section{Improvement in the spectrum}
We use the previously described meson and glueball operators to extract the iso-vector and iso-scalar scalar spectrum in our $N_f = 2$ ensemble. To showcase the usefulness of our operator construction, we compare the results with those coming from unimproved operators in both meson and glueball cases. 

\subsection{Iso-vector channel}
\begin{figure}
    \centering
    \includegraphics[width=0.49\textwidth]{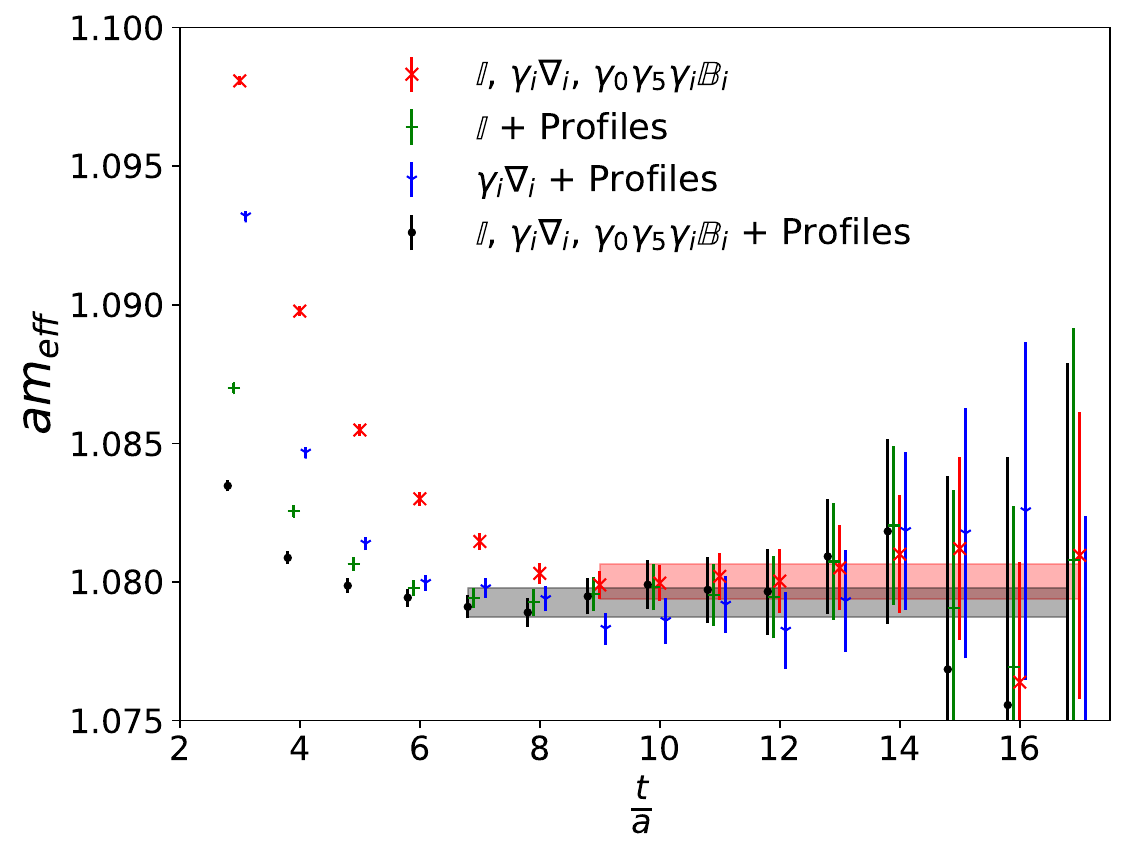}
    \caption{Ground state effective masses coming from separate GEVPs. Red points come from a $3\times 3$ GEVP using $\Gamma = \mathbb{I}, \gamma_i \nabla_i, \gamma_0 \gamma_5 \gamma_i \mathbb{B}_i$. The red band is the corresponding plateau average. Blue and green points come from solving a $7\times 7$ GEVP for $\mathbb{I}$ and $\gamma_i \nabla_i$ respectively using distillation profiles. Black points come from building a $21\times 21$ correlation matrix including all three choices of $\Gamma$ and 7 profiles for each one and then pruning it down to an $8\times 8$ matrix to solve the GEVP. The black band is the corresponding plateau average. We omit the results from the $7\times 7$ GEVP using only $\gamma_0 \gamma_5 \gamma_i \mathbb{B}_i$ since the statistical uncertainties are much larger than for the other cases.}
    \label{fig:Masses_Comparison_IV}
\end{figure}

For the iso-vector $(I=1)$ channel, we only use meson operators projected to this flavor quantum number. In practical terms, this results in requiring only quark-connected correlations. The correlation matrix has entries given by
\begin{align}
    C_{ij}(t) &= -\text{Tr}\left(  \Phi_i[t] \tau[t,0] \bar{\Phi}_j[0] \tau[0,t]  \right),
    \label{eqn:Connected}
\end{align}
where $\Phi_i[t]$ is the elemental for a fixed choice of $\Gamma$ and profile at time $t$ and $\bar{\Phi}_i[t]$ is the analogous elemental with $\bar{\Gamma} = \gamma_0 \Gamma^{\dagger} \gamma_0$. To showcase the effectiveness of our operator construction, we extract the ground state effective masses using three different types of GEVPs. First, solve a $3\times 3$ GEVP using $\Gamma = \mathbb{I}, \gamma_i \nabla_i, \gamma_0 \gamma_5 \gamma_i \mathbb{B}_i$ and standard distillation. Second, we solve three separate $7 \times 7$ GEVPs; one for each choice of $\Gamma$. This is the approach first presented in \cite{Urrea-2022}. Finally, we solve a $21 \times 21$ GEVP including the three choices of $\Gamma$ and 7 profiles for each one. For this last GEVP the basis size is considerably large, so we use the same pruning procedure used in Ref.~\cite{Urrea-2022} taken from \cite{Balog-1999, Niedermayer-2001, Blossier}, i.e. projecting the correlation matrix at all times onto the singular vectors corresponding to the largest singular values of a fixed reference time. In doing so, we reduce the basis size preserving a set of orthonormal operators which retain the information of the low-lying states while at the same time removing possible numerical instabilities. This approach mixes operators with different $\Gamma$ and therefore hinders the spin-identification strategy of \cite{Dudek-2010}. Since in this work we are not interested in identifying continuum quantum numbers, this issue does not trouble us. However, one way to circumvent this would be a partial pruning as presented in \cite{Urrea-Scalar}. By selecting the significant singular vectors of the diagonal sub-matrices of a fixed $\Gamma$ and embedding them into vectors in the full space of all operators we can reduce the basis size and sensitivity to statistical noise without mixing different $\Gamma$ operators. In all cases we normalize the correlation matrix before any pruning using
\begin{align*}
    C_{ij}(t) \rightarrow \frac{C_{ij}(t)}{\sqrt{C_{ii}(t_R) C_{jj}(t_R)}},
\end{align*}
where we use $t_R = 2a$ unless stated otherwise and we solve the GEVPs using as reference time $t_G = t_R$, i.e. we solve the equation
\begin{align}
    C(t) \mathbf{v}_i(t,t_G) &= \lambda_i(t,t_G) C(t_G) \mathbf{v}_i(t,t_G).
\end{align}
For the pruning we use the 8 largest singular values, unless stated otherwise, at time $t_R$. In Fig. \ref{fig:Masses_Comparison_IV} we show the ground state effective masses calculated in the different GEVPs previously. We point out two major observations. First, the approach to the plateau is much faster when using a single $\Gamma$ with profiles, either $\mathbb{I}$ or $\gamma_i \nabla_i$, compared to combining all three choices of $\Gamma$ with standard distillation. As was shown when distillation was first presented \cite{Peardon-2009} as well as in recent work on static-light spectroscopy \cite{Struckmeier-2025}, one can find an optimum number of distillation vectors for the standard approach. However, this requires tuning via trial and error and is unfeasible to optimize the method this way for every state of interest. Furthermore, Ref.~\cite{Struckmeier-2025} shows the use of optimal profiles always outperforms the standard approach. In this sense, it very well might be that $200$ vectors is not the optimal choice for this ground state and the red points could converge faster to a plateau after tuning. Fortunately, the use of optimal distillation profiles eliminates the need for this time-consuming, state-dependent tuning by allowing the GEVP to find the optimal profile in each case. Since distillation-based studies often fix the same number of vectors for all channels and it is not possible to predict the optimal choice, the red points in Fig. \ref{fig:Masses_Comparison_IV} represent the results we would obtain in this approach and the use of optimal profiles brings a major improvement. Second, the combination of profiles with not one but all available choices of $\Gamma$ further accelerates the convergence of the effective masses to a plateau. While going from standard distillation to using optimal profiles brings the largest gain, extending the basis with multiple $\Gamma$'s is still useful for the ground state. We extract mass plateaus comparing the cases of the three $\Gamma$ operators with and without distillation profiles, given by $am = 1.07927(54)$ and $am = 1.08002(63)$ respectively. The earlier plateau coming from the improvement results in smaller statistical uncertainty for the plateau average \footnote{We extract the plateau value as a weighted average using the inverse error squared as weights.}.
\par
\begin{figure}
    \centering
    \includegraphics[width=0.4\textwidth]{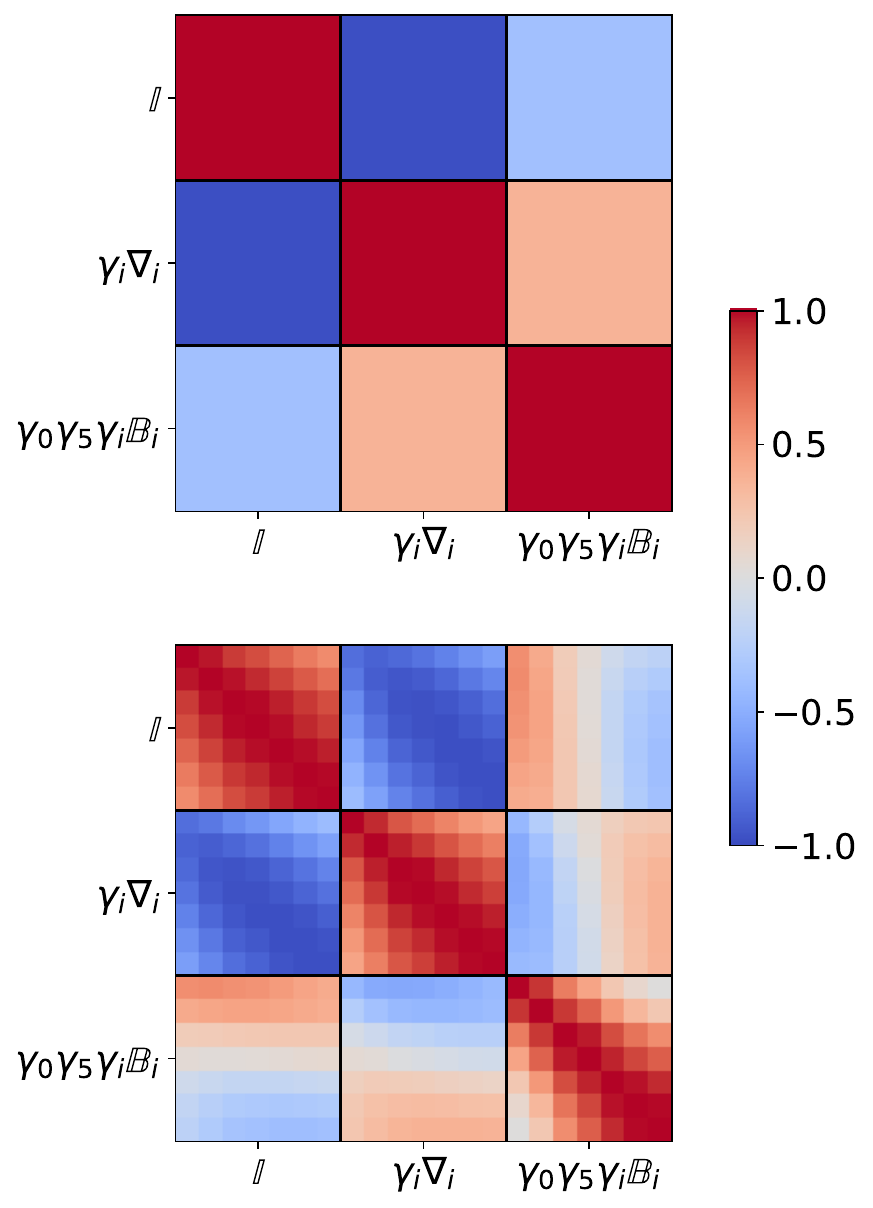}
    \caption{Correlation matrices at $t = 2a$ using three choices of $\Gamma$ with standard distillation (up) and 7 profiles per $\Gamma$ (down).}
    \label{fig:C_IV_Ref}
\end{figure}
\begin{figure*}
    \centering
    \includegraphics[width=0.8\textwidth]{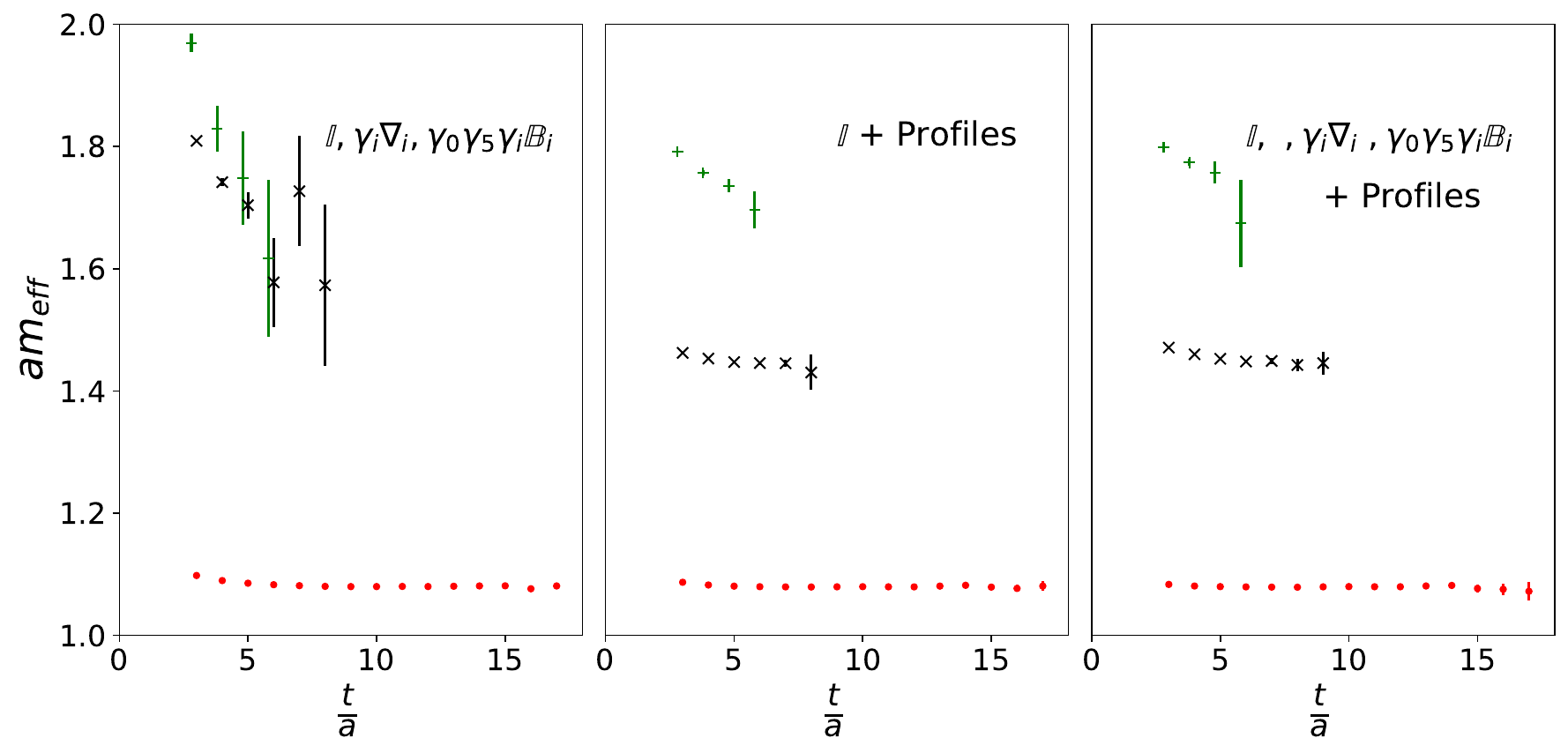}
    \caption{Low-lying $A_1^{++}$ iso-vector spectrum determined via different GEVPs. \textbf{Left:} Using $\Gamma = \mathbb{I}, \gamma_i \nabla_i$, $\gamma_0 \gamma_5 \gamma_i \mathbb{B}_i$ with standard distillation. \textbf{Middle:} Using $\Gamma = \mathbb{I}$ with 7 profiles. \textbf{Right:} Using $\Gamma = \mathbb{I}, \gamma_i \nabla_i$, $\gamma_0 \gamma_5 \gamma_i \mathbb{B}_i$ with 7 profiles for each $\Gamma$.}
    \label{fig:States_IV}
\end{figure*}

In Fig. \ref{fig:C_IV_Ref} we show the correlation matrices at $t=2a$ including the different choices of $\Gamma$ both with standard distillation and with 7 profiles per choice of $\Gamma$, the former being also contained in the latter. Each entry in the $3\times 3$ matrix of standard distillation gets extended into a $7\times 7$ block thanks to the inclusion of profiles. These new blocks do not introduce numerically-unstable degeneracies since the operators coming from different profiles are still sufficiently different, which is graphically evidenced by the different shades contained in a same block. This is particularly clear in the blocks relating $\gamma_0 \gamma_5 \gamma_i \mathbb{B}_i$ with the other two operators: different profile combinations lead to correlations with different signs. Even after the inclusion of profiles, the block-like coupling between the different $\Gamma$'s keeps the same qualitative behavior: a large correlation in absolute value between $\mathbb{I}$ and $ \gamma_i \nabla_i$ but a small correlation in absolute value between these two and $\gamma_0 \gamma_5 \gamma_i \mathbb{B}_i$. To further emphasize the usefulness of introducing the profiles to the basis containing different choices of $\Gamma$, we can look at the resulting spectrum with and without using them. Fig. \ref{fig:States_IV} shows the lowest three states determined via three different GEVPs. On the left, we use $\Gamma = \mathbb{I}, \gamma_i \nabla_i, \gamma_0 \gamma_5 \gamma_i \mathbb{B}_i$ with standard distillation. In the middle, we use $\Gamma = \mathbb{I}$ with 7 different profiles. On the right, we use all three choices of $\Gamma$ with 7 profiles for each one. While Fig.~\ref{fig:Masses_Comparison_IV} shows all three approaches resolve a consistent ground state, the benefit of extending our basis is already seen on the first excitation state in Fig.~\ref{fig:States_IV}. When using only the three choices of $\Gamma$ with standard distillation, this first excitation cannot be clearly resolved and neither can the following one, so out of three possible states only one is resolved. 

\subsection{Iso-scalar channel}
For the iso-scalar ($I=0$) case, we have both meson and glueball operators with this flavor quantum number. These different types of operators can mix non-trivially in the off-diagonal entries of a correlation matrix and therefore we must take them all into account for extracting the spectrum. The full correlation matrix is given block-wise by
\begin{align}
    C(t) &= \begin{pmatrix}
        C_{MM}(t) + 2D_{MM}(t) & -\sqrt{2}C_{MG}(t) \\
        -\sqrt{2} C_{GM}(t) & C_{GG}(t)
    \end{pmatrix}
\end{align}
where $C_{MM}(t)$ is given by Eq. \ref{eqn:Connected} and the remaining blocks are
\begin{align}
    D_{MM}(t) &= \text{Tr}\left( \Phi[t] \tau[t,t]  \right) \text{Tr}\left( \bar{\Phi}[0] \tau[0,0] \right) \nonumber\\
    C_{MG}(t) &= \text{Tr}\left( \Phi[t] \tau[t,t]  \right) G[0] \nonumber\\
    C_{GM}(t) &= G[t] \text{Tr} \left( \bar{\Phi}[0] \tau[0,0] \right) \nonumber\\
    C_{GG} &= G[t]G[0],
\end{align}
where $G[t]$ is a glueball operator at time $t$. Each block contains the correlations mixing the different operators involved, e.g. $C_{MG}(t)$ has as many rows as there are meson operators and as many columns as there are glueball operators. To display the advantages of our operator construction, we will first use meson and glueball operators separately and compare the results with the current approaches. Afterwards, we combine our improved operators to extract the iso-scalar spectrum by either including all the correlation matrix or only some of its diagonal blocks in the GEVP.

\subsubsection{Glueball operators only}
We list the $A_1^{++}$ glueball operators employed up to mass-dimension 6 omitting overall normalization factors:
\begin{align}
    \mathcal{O}_1(t) &= \sum_{\vec{x},i} \text{Tr}\left( \mathcal{B}_i(\vec{x},t)^2  \right) \nonumber\\
    \mathcal{O}_2(t) &= \sum_{\vec{x}} \epsilon_{ijk} \text{Tr}\left( \mathcal{B}_i(\vec{x},t) \mathcal{B}_j(\vec{x},t) \mathcal{B}_k(\vec{x},t)  \right) \nonumber\\
    \mathcal{O}_3(t) &= \sum_{\vec{x}} \text{Tr} \left(  \mathcal{B}_i(\vec{x},t) \sum_{k} D_k^2 \mathcal{B}_i(\vec{x},t) \right) \nonumber\\
    \mathcal{O}_4(t) &= \sum_{\vec{x}} \text{Tr}\left(  \mathcal{B}_1(\vec{x},t) \left( 2D_1^2 - D_2^2 - D_3^2  \right) \mathcal{B}_1(\vec{x},t) \right) \nonumber\\
    &+ \text{Tr}\left(  \mathcal{B}_2(\vec{x},t) \left( 2D_2^2 - D_3^2 - D_1^2  \right) \mathcal{B}_2(\vec{x},t) \right)\nonumber\\
    &+ \text{Tr}\left(  \mathcal{B}_3(\vec{x},t) \left( 2D_3^2 - D_1^2 - D_2^2  \right) \mathcal{B}_3(\vec{x},t) \right)\nonumber\\
    \mathcal{O}_5(t) &= \sum_{\vec{x}} |\epsilon_{ijk}||\epsilon_{jmn}| \text{Tr}\left( \mathcal{B}_i(\vec{x},t) D_m D_n \mathcal{B}_k(\vec{x},t)  \right).
    \label{eqn:GlueballOperators}
\end{align}
Although we build them using Clebsch-Gordan coefficients from the cubic group, we can still make a qualitative statement about the contributions they receive from continuum $J$. By counting the number of fields included, we expect the first two operators to have a leading contribution from $J=0$, since we need at least 4 fields to get $1\otimes 1 \otimes 1 \otimes 1 \rightarrow 4$. We also expect the third operator to have a leading contribution from $J=0$ since we have coupled the derivatives to $A_1$, equivalent to how we can couple them as $1 \otimes 1 \rightarrow 0$. This results in having $1 \otimes 0 \otimes 1 \rightarrow 0$ for the overall operator. For the last two operators, the derivatives are not coupled as $1 \otimes 1 \rightarrow 0$. In our case, we have chosen to couple them in the lattice irreps $E$ and $T_2$, which are related to the continuum $J=2$. We have $1 \otimes 2 \otimes 1$, which has contributions from $J = 0$ and $J = 4$. Since we have chosen not to isolate these contributions but rather work directly in our choice of cubic group irreps, we cannot rule out leading contributions from $J=4$ as well as $J = 0$ in the last two operators.  
\par
\begin{figure}
    \centering
    \includegraphics[width=0.49\textwidth]{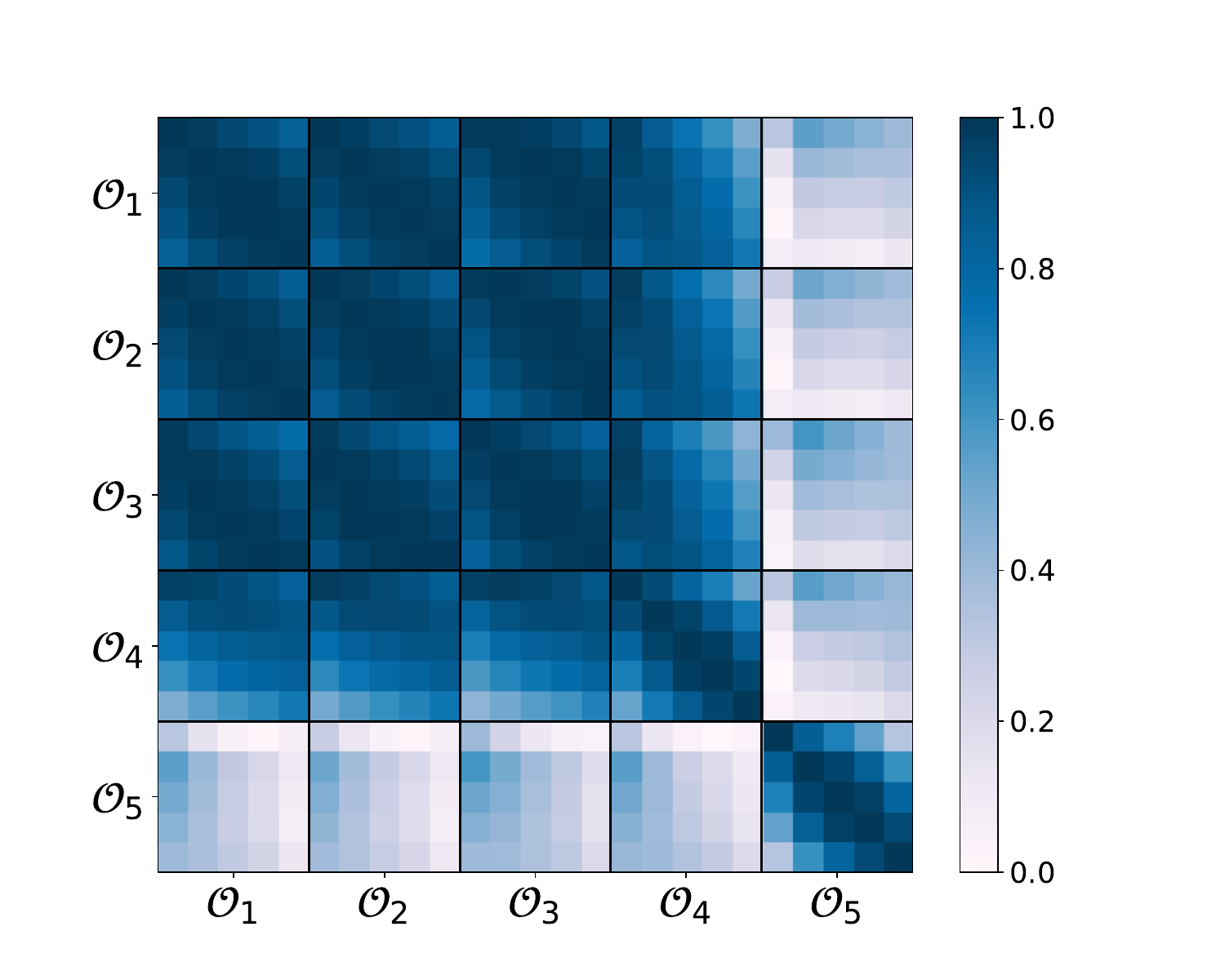}
    \caption{Normalized correlation matrix taking the absolute value of the entries at $t=a$ using the alternative glueball operators at different levels of APE smearing. The operators $\mathcal{O}_i$ are defined in Eq. \ref{eqn:GlueballOperators}.}
    \label{fig:FullBasis_Glueballs}
\end{figure}
\begin{figure}
    \centering
    \includegraphics[width=0.49\textwidth]{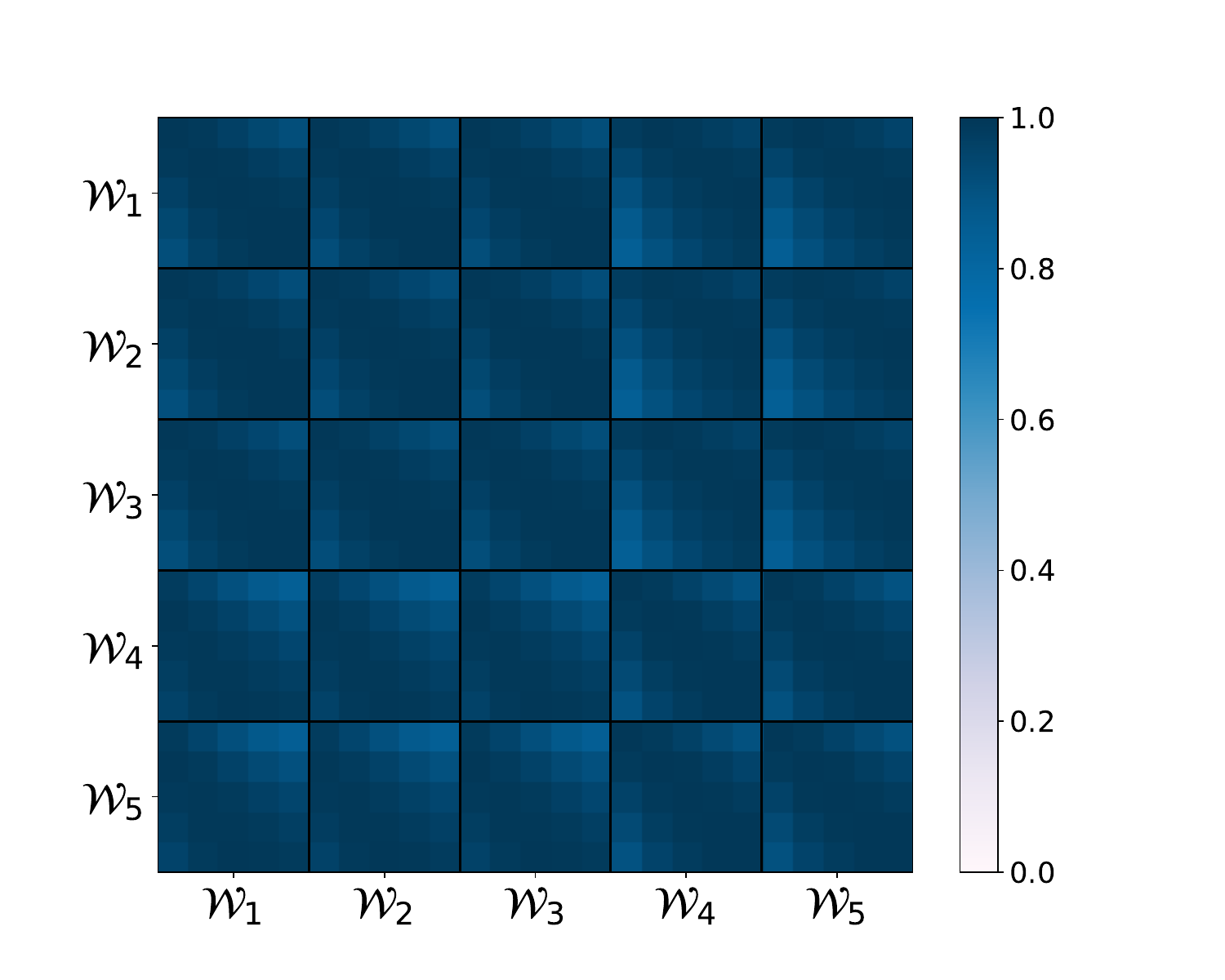}
    \caption{Normalized correlation matrix taking the absolute value of the entries at $t=a$ using 5 different Wilson loop shapes at different levels of APE smearing.}
    \label{fig:FullBasis_Loops}
\end{figure}

We test this approach first by looking at the resulting correlation matrix. We take not only the above five operators but also evaluate them at 5 different levels of APE smearing \cite{Albanese-1987} with $\alpha = 0.35$ and $5,10,15,20,30$ iterations, yielding a total of 25 operators. Fig. \ref{fig:FullBasis_Glueballs} shows the normalized correlation matrix at $t=a$ taking the absolute value of the entries. The approximately block-diagonal form confirms our expectation about the leading $J=0$ contributions to the first three operators, which are grouped together. The fourth operator couples to the first three, hinting at a strong $J=0$ content despite a possible $J=4$ we did not explicitly suppress. It also seems more sensitive to the different levels of smearing when coupling to itself and the other four operators. This is advantageous, since it means the operator is less prone to smearing-induced degeneracy. The last operator is rather decoupled from the others, hinting at a strong $J=4$ contribution, as well as less prone to smearing-induced degeneracy. We remark we did not build all our operators to have an assigned leading $J$. Rather, we construct them to be as linearly independent as possible based on their construction from the chromo-magnetic and derivative fields. Nonetheless, building them to have an assigned $J$ would simply require to use $SO(3)$ Clebsch-Gordan coefficients for the tensor products of representations and the corresponding subduction coefficients. We note that, as in the case of meson operators with fixed $\Gamma$ and different profiles, fixing the glueball operator construction and changing the level of APE smearing does not change the qualitative behavior of the correlation matrix. Interestingly, the entries change sign when relating the fifth operator with the first three depending on the level of smearing used. This hints at the importance of smearing even when using these alternative operators. For comparison with the traditional approach, we have taken 5 different loop shapes from \cite{Berg-1983} and measured them at 5 different levels of APE smearing with $\alpha = 0.35$ and $10,15,20,25$ iterations. Fig. \ref{fig:FullBasis_Loops} shows the normalized correlation matrix at $t=a$, with $\mathcal{W}_i$ labeling the different loop shapes. Unlike for the operators based on the $\mathcal{B}_i$ and $D_i$, the correlation matrix is not close at all to being block-diagonalized and remains dense. Furthermore, the near-degeneracies observed in previous studies, e.g. in \cite{Sakai-2023}, are clear here; most entries being close to $1$. 

\begin{figure}
    \centering
    \includegraphics[width=0.45\textwidth]{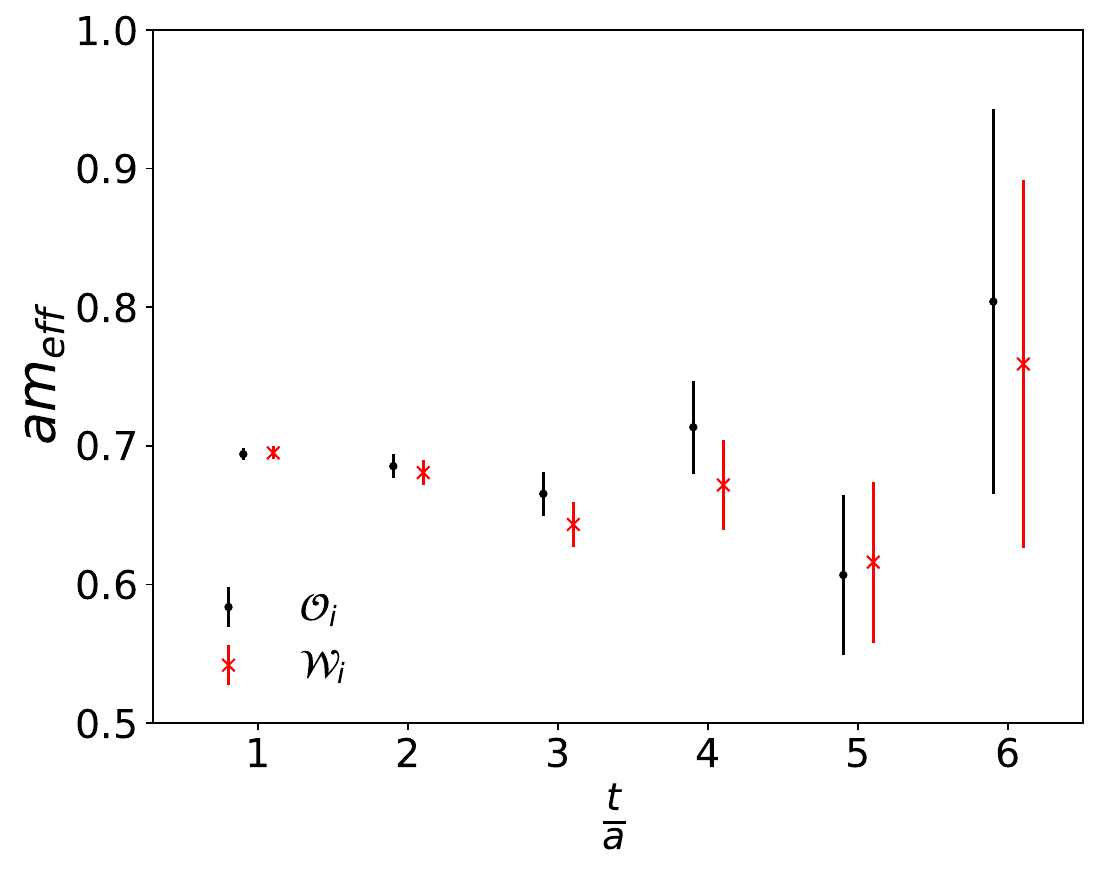}
    \caption{Ground state effective masses from two separate GEVPs: one using only the Wilson loop operators $\mathcal{W}_i$ and another one using the alternative operators $\mathcal{O}_i$.}
    \label{fig:Masses_Comparison_Glueballs}
\end{figure}

We have already highlighted some important aspects of the operators $\mathcal{O}_i$ built from the fields $\mathcal{B}_i$ and $D_i$; they follow a physical intuition in their construction including a systematic procedure to include an arbitrary number of these fields, their implementation is much more straightforward and suitable for parallelized calculations and the simplicity of their construction extends to the cases of non-zero total spatial momentum. These three features make them preferable compared to the commonly-used spatial Wilson loops. Nonetheless, these two approaches are not completely independent. Just as we chose to discretize the fields to recover them up to a fixed power of the lattice spacing via a clover definition for $\mathcal{B}_i$ and a symmetric difference for $D_i$, we could also find a suitable linear combination of spatial Wilson loops which recover a given operator involving any of these fields up to a given power of the lattice spacing. While this is a well-known procedure extensively applied mainly for the construction of improved gauge actions, e.g. \cite{Luscher1985, Luscher1985-Erratum, Gunther2016-av}, and has also been used for glueball operators involving the 4-dimensional $F_{\mu \nu}$ \cite{Sun-2018}, it entails the limitations related to implementing arbitrarily complicated loop shapes as well as the parallelization of the calculations. If we use the appropriate loop shapes $\mathcal{W}_i$, we can find a linear transformation relating the operators in terms of those loop shapes with the same continuum-like operators (up to a power of the lattice spacing) which the operators $\mathcal{O}_i$ also recover. Since here we have taken only five loop shapes, we do not expect an exact matching to happen. However, we do expect both sets of operators to span a very similar space. The advantage in the choice of basis for the $\mathcal{O}_i$ lies on how well they individually couple to continuum-like states as well as their linear independence. The former property is present in the HadSpec operators for mesons; one could choose a set of operators containing the same number of derivatives not built from a subduction from $SO(3)$ such that the there is no memory from the continuum $J$, however as long as the same space of operators is spanned, the spectrum will remain the same thanks to the GEVP being invariant under unitary transformations. In Fig. \ref{fig:Masses_Comparison_Glueballs} we show the ground state effective masses coming from two separate GEVPs: one with the $\mathcal{W}_i$ operators and another with the $\mathcal{O}_i$ operators, both extended with the different levels of APE smearing. The near-degeneracy between the effective masses confirms both sets of operators span almost the same space. We emphasize that the human and computational effort to obtain the $\mathcal{O}_i$ results is much smaller than for the $\mathcal{W}_i$ ones. 

\subsubsection{Meson operators only}
\begin{figure}
    \centering
    \includegraphics[width=0.49\textwidth]{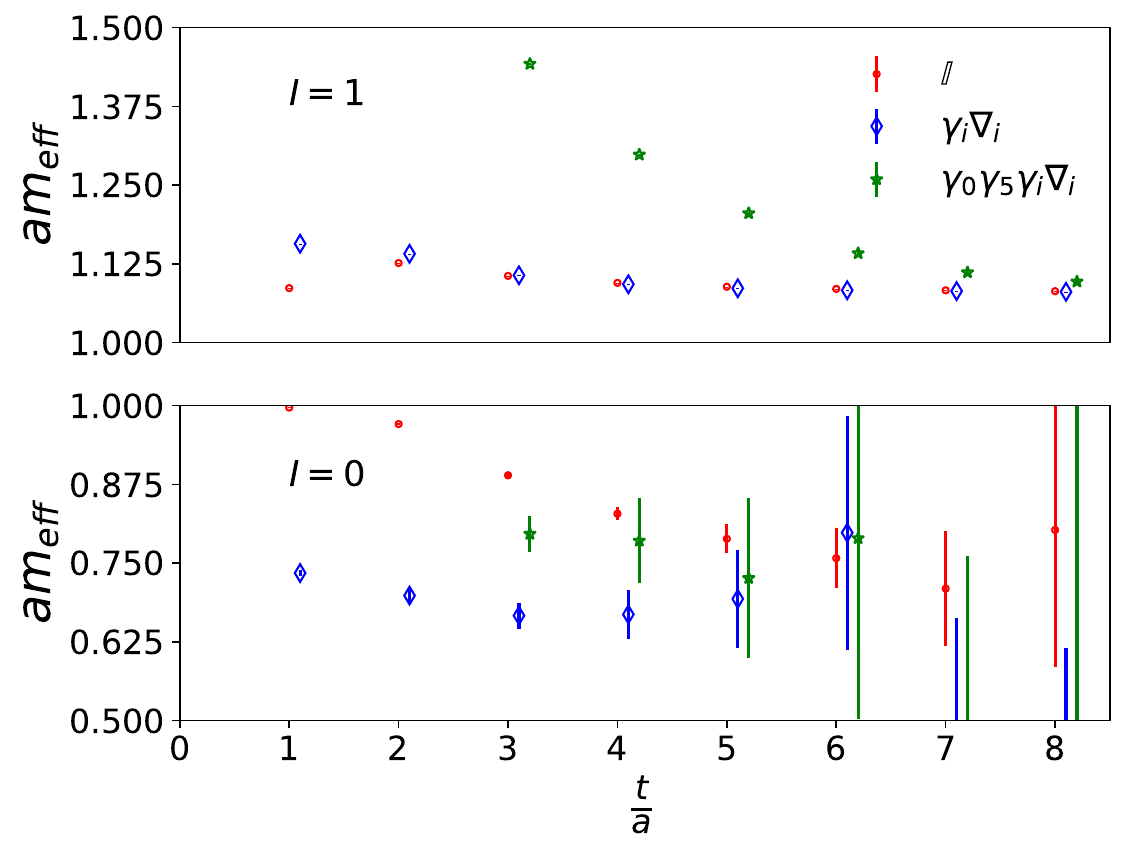}
    \caption{Ground state effective masses using each choice of $\Gamma$ and standard distillation for the iso-vector $(I = 1)$ channel (above) and the iso-scalar $(I=0)$ channel (below).}
    \label{fig:Masses_Comparison_Std_IS}
\end{figure}

We now calculate the spectrum using only iso-scalar meson operators with the same $\Gamma$ structures and profiles as in the iso-vector case. The corresponding correlation points include disconnected contributions, well-known to have larger statistical uncertainty compared to the connected contributions. Since the statistical errors in these iso-scalar correlations grow exponentially with the temporal separation, it is of fundamental importance to be able to extract reliable spectrum information at short time separations. For this reason, optimizing the meson operators involved is a key improvement. A first aspect to take into account is the choice of $\Gamma$. One might think increasing the number of derivatives increases the contributions from excitations, which is a key ingredient for the GEVP, and so if we had to choose only one operator, we would go with the simplest one available, i.e. $\Gamma = \mathbb{I}$ for the $A_1^{++}$ channel. In Fig. \ref{fig:Masses_Comparison_Std_IS} we show the ground state effective masses obtained using with standard distillation and each choice of $\Gamma$ separately for the iso-vector case in the upper panel and for the iso-scalar one in the lower panel. We restrict the time separations to those where the iso-scalar case has signal before the errorbars become too large. In the iso-vector channel the effective masses follow the above-mentioned intuition rather well; $\Gamma = \mathbb{I}$ has the fastest approach to a plateau, followed very closely by $\Gamma = \gamma_i \nabla_i$. Apart from a very small difference, these two operators behave similarly well. The case $\Gamma = \gamma_0 \gamma_5 \gamma_i \mathbb{B}_i$ has a much slower convergence and therefore it is mostly useful as part of a GEVP to determine radial excitations. In the iso-scalar channel the behavior is very different. The fastest approach to a plateau is shown by $\Gamma = \gamma_i \nabla_i$, whose points remain significantly below those of $\Gamma = \mathbb{I}$ until around $t=4a$ before they become consistent with each other. This makes the $\Gamma = \gamma_i \nabla_i$ a better choice for this channel, a conclusion we could not have reached by looking only at the iso-vector data. The errorbars for this operator are nonetheless still larger than for the no-derivative one, so combining them together to get both their benefits remains the best path to follow. With $\Gamma = \gamma_0 \gamma_5 \gamma_i \mathbb{B}_i$ we observe even larger errors, however the approach to the plateau is comparable to that of $\Gamma = \mathbb{I}$, contrary to the iso-vector case where it had the slowest convergence to a plateau by a significant margin. This rather major change in behavior between different flavor symmetry channels is yet another reason why a careful construction of operator basis is important, as those operators which appear optimal for one channel might not be so for another one.

\begin{figure}
    \centering
    \includegraphics[width=0.49\textwidth]{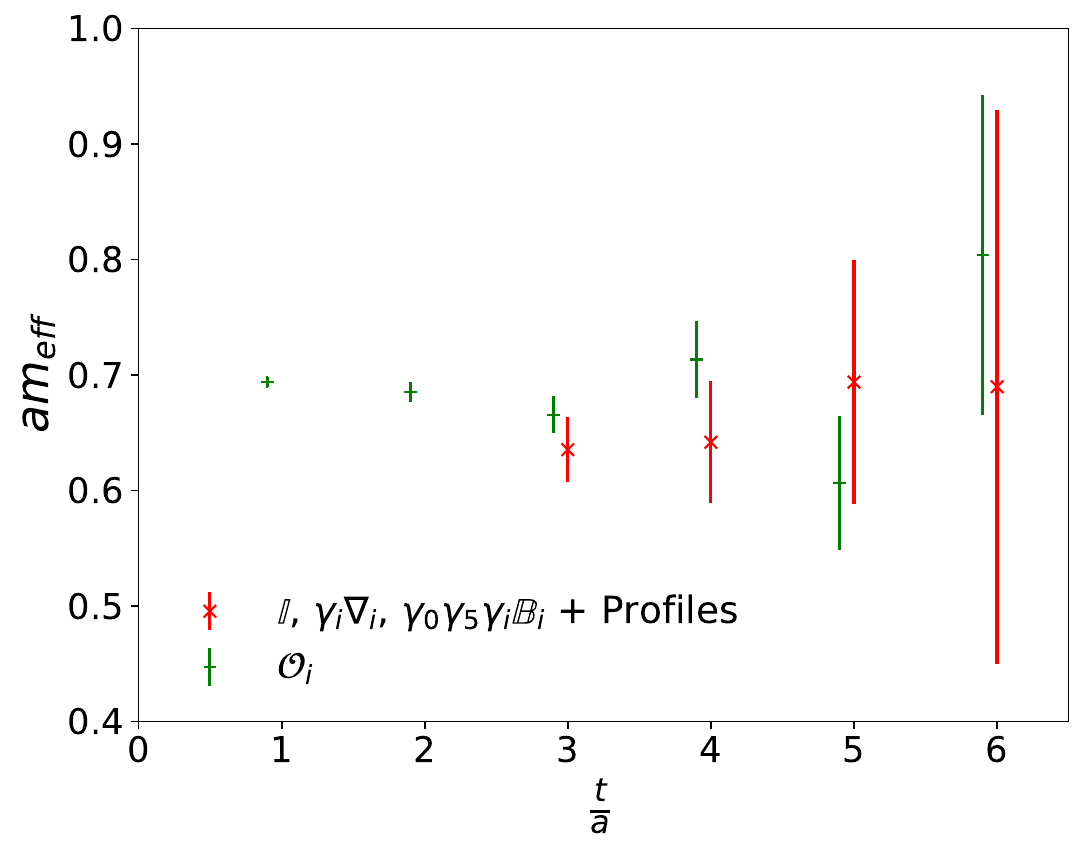}
    \caption{Ground state effective masses from the meson operator basis including $\Gamma = \mathbb{I}, \gamma_i \nabla_i, \gamma_0 \gamma_5 \gamma_i \mathbb{B}_i$ with 7 profiles each (red) as well as from a basis containing the alternative glueball operators $\mathcal{O}_i$ (green). Note the meson results use $t_G = 2a$ while the $\mathcal{O}_i$ use $t_G = 0$.}
    \label{fig:Masses_Comparison_IS_G&M}
\end{figure}

As with the iso-vector case, we build a GEVP including all choices of $\Gamma$ and $7$ profiles for each one, pruning the $21 \times 21$ correlation matrix down to an $8\times 8$. In Fig. \ref{fig:Masses_Comparison_IS_G&M} we present the resulting ground state effective masses using meson-only operators, and we also include the same masses obtained for the glueball-only operators $\mathcal{O}_i$ from Fig. \ref{fig:Masses_Comparison_Glueballs}. We note here the two GEVPs have different $t_G$; $0$ for the $\mathcal{O}_i$ and $2a$ for the meson operators. Regardless, the effective masses are completely consistent with each other and reveal an $A_1^{++}$ iso-scalar ground state considerably lighter than the $A_1^{++}$ iso-vector ground state.
In \cite{Urrea-2022} we determined the lowest-lying iso-vector state, the $A_1^{-+}$ ("pion"), to have a mass in lattice units equal to $am = 0.74986(8)$, which makes the iso-scalar $A_1^{++}$ state we see in Fig. \ref{fig:Masses_Comparison_IS_G&M} stable in our setup. With a mass in lattice units slightly above $am=0.6$, this would correspond to $\approx$ $1900$ MeV, consistent with quenched predictions of the $A_1^{++}$ ground state glueball, e.g. \cite{Peardon-1999}. While this is already a promising hint of the nature of this state, we can obtain additional information by building a correlation function which includes both meson and glueball operators which until now we have analyzed separately. 
\subsubsection{Meson and glueball operator mixing}
\begin{figure}
    \centering
    \includegraphics[width=0.55\textwidth]{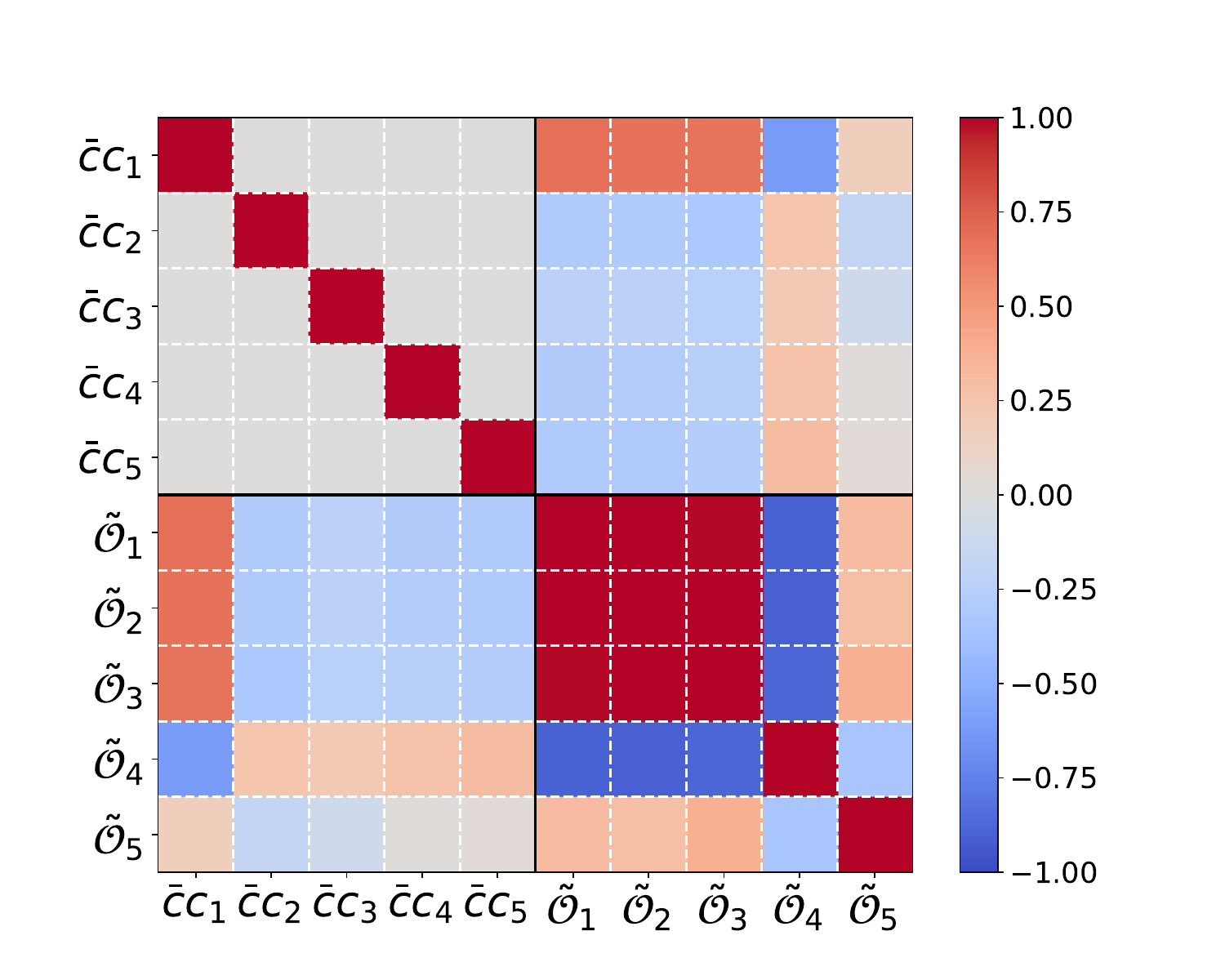}
    \caption{Normalized correlation matrix at $t=a$ including the pruned charmonium operators $\hat{c}c_i$ and the pruned glueball operators $\tilde{\mathcal{O}}_i$. See the text for our choice of pruning for the construction of the meson and glueball operators. In particular, the charmonium operators are chosen to exactly diagonalize the first $5\times 5$ block at $t=a$.}
    \label{fig:FullBasis_Mixing}
\end{figure}

\begin{figure}
    \centering
    \includegraphics[width=0.45\textwidth]{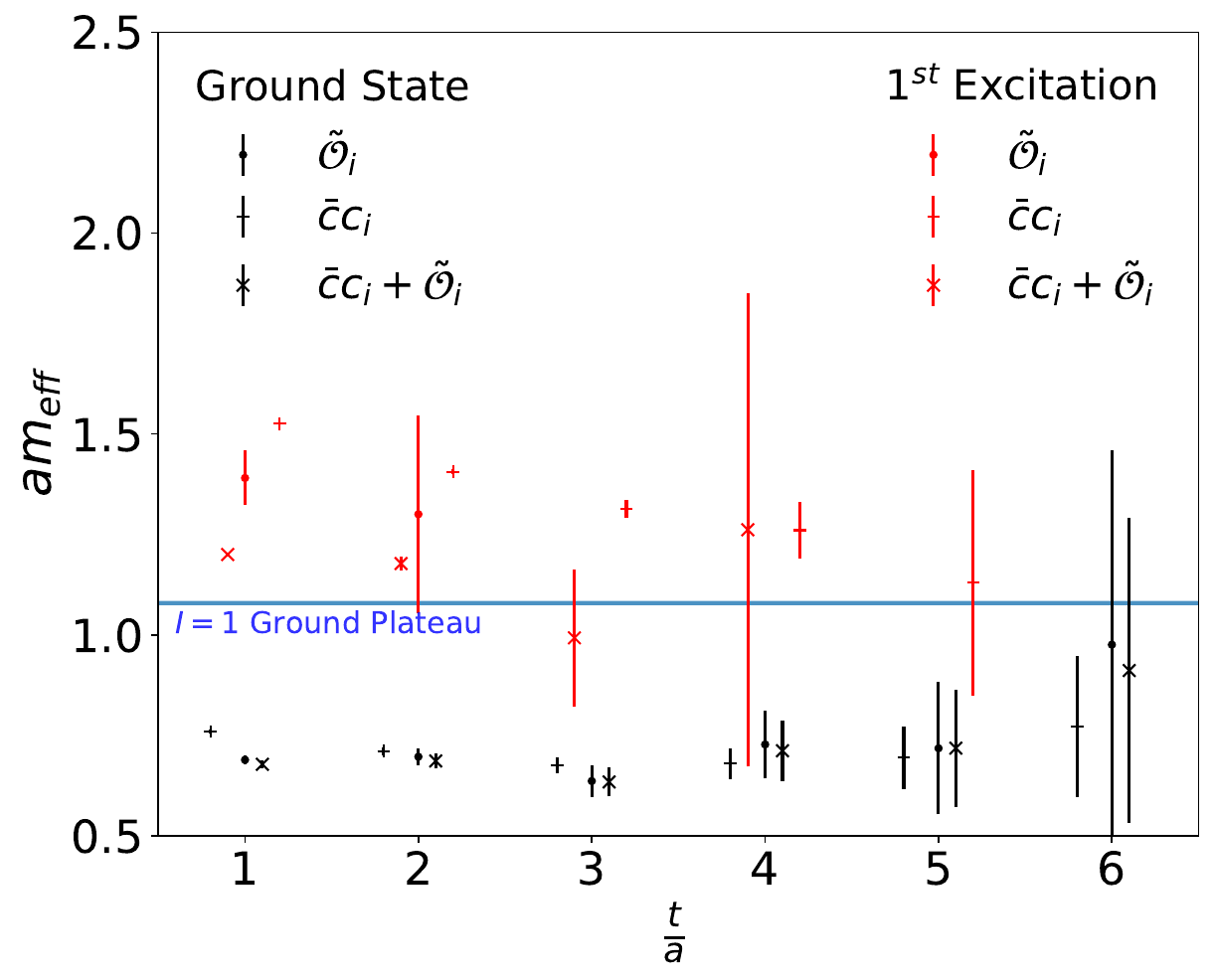}
    \caption{Iso-scalar ground and first excited state effective masses obtained from three different GEVPs: glueball operators only ($\tilde{\mathcal{O}}_i$), charmonium operators only ($\bar{c}c_i$) and both types combined. The blue band corresponds to the plateau average of the iso-vector case for comparison.}
    \label{fig:Masses_Comparison_All}
\end{figure}

To put together meson and glueball operators in our basis, we first need to account for a few issues. First, we have $21$ meson operators and $25$ glueball operators, which would result in a $46 \times 46$ correlation and numerical problems related to stability will probably arise. Just as we pruned when analyzing the operators separately, we will also prune this correlation matrix. However, since we wish to quantify how relevant the two types of operators used are to resolve the energy eigenstates, we cannot prune using the singular vectors of the full matrix, as these get non-zero entries for all operators. Instead, we use the partial pruning approach we also used in \cite{Urrea-2025} in the context of isolating light-meson and charmonium operators; we calculate pruning vectors for the meson and glueball operators separately. This approach improves the numerical stability of the GEVP while still retaining the distinction between glueball and meson operators. Second, we have to be careful when choosing the reference time $t_G$ for the GEVP. While the choice $t_G = 0$ worked well for the glueball operators alone, the correlation matrix for the meson operators at this time is not always positive definite owing to operator-dependent contact term effects. Since we can only study the mixing between meson and glueball operators using the gauge configurations where we have perambulators, the reduced statistics will increase the uncertainty in the glueball correlation functions compared to Fig. \ref{fig:FullBasis_Glueballs} and we should therefore try to work with the smallest $t_G$ possible. The correlation matrix using only $\Gamma = \gamma_i \nabla_i$ with 7 profiles is positive-definite at $t=0$ and this operator had already the fastest convergence to a plateau when using standard distillation, so we use this one with different profiles to mix with the glueball operators. We choose as pruning vectors the $5$ ones corresponding to the largest singular values of the corresponding $7\times 7$ block. For the glueball operators we extract the normalized diagonal blocks corresponding to each operator $\mathcal{O}_i$ including different levels of smearing and calculate the singular vector corresponding to the largest singular value. This results in $5$ pruning vectors which still distinguish between the different choices of $\mathcal{O}_i$ and simply linearly combine the same type of operator at different smearing levels. This preserves the almost block-diagonal structure of the glueball block in our correlation matrix. Fig. \ref{fig:FullBasis_Mixing} shows the resulting correlation matrix at $t=a$, which we again normalize for clear visualization. We label as $\bar{c}c_i$, $i=1,...,5$ the five different meson operators built from the pruning vectors and $\tilde{\mathcal{O}}_i$, $i=1,...,5$ are the five different glueball operators built from the corresponding pruning vectors which still conserve the structure of each original $\mathcal{O}_i$. As expected, the glueball $5\times 5$ block preserves its almost block-diagonal structure as in Fig. \ref{fig:FullBasis_Glueballs}. The meson block is the identity matrix at $t=a$ due to how we chose the pruning and normalization, however this no longer holds at $t > a$. The off-diagonal blocks corresponding to the mixing between meson and glueball operators have entries with very small magnitude compared to the diagonal blocks, showing how while there is definitely non-trivial mixing between these two types of operators, it is not as strong as most of the mixing among a same type of operator.  
We extract the spectrum by solving three different GEVPs: one with the full $10\times 10$ matrix, another one using only the $5\times 5$ charmonium operator block and another one using the $5\times 5$ glueball operator block. This way we can compare the quality of the resulting effective masses in terms of convergence to a plateau as well as statistical uncertainty. Fig. \ref{fig:Masses_Comparison_All} shows the resulting effective masses for the ground state and first radial excitation from the three different GEVPs. All three display a consistent ground state, with the charmonium-only case having a marginally slower convergence yet smaller uncertainties. For the case of the first radial excitation there is a more varied behavior. The clearest signal in terms of uncertainties comes from the charmonium-only case, which is expected, yet its convergence remains the slowest. It is only when we mix these with the glueball operators that a faster convergence is achieved, and the glueball operators give only very little signal before the errors become too large. While the mixing leads to a significant improvement in the first radial excitation, one could argue that for the ground state the glueball operators were already good enough. This however we cannot know a priori and performing the mixing, particularly to resolve a first excitation, remains the systematically correct approach.
\par
\begin{figure}
    \centering
    \includegraphics[width=0.45\textwidth]{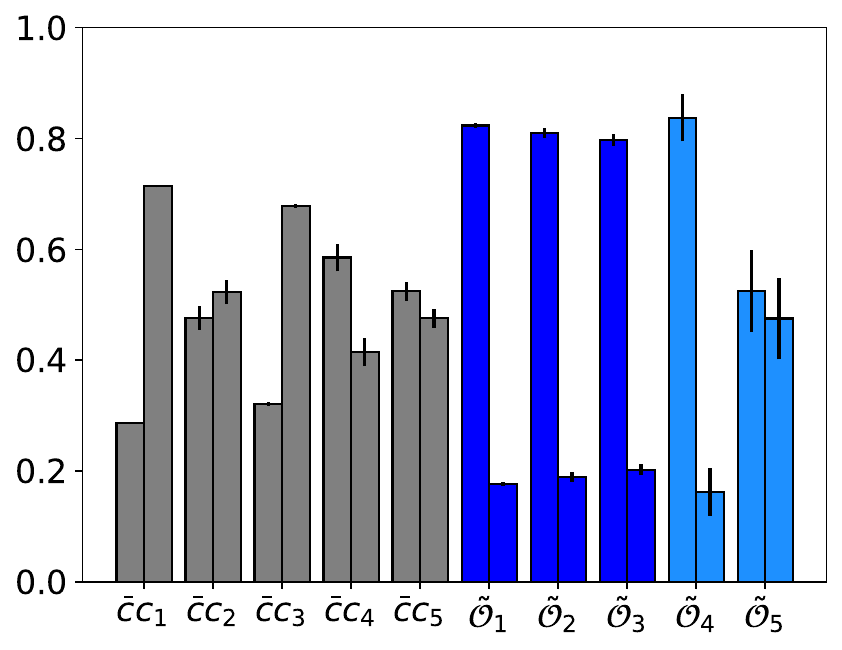}
    \caption{Overlaps between states created by the charmonium $\bar{c}c_i$ (gray) and glueball $\tilde{\mathcal{O}}_i$ (shades of blue) operators and the energy energy eigenstates. The first column per operator is the ground state, the second one is the first excitation. Within the glueball operators, we distinguish the ones with dominant $J=0$ contributions (dark blue) from those which a priori can have also sizable $J > 0$ contributions (light blue).}
    \label{fig:Overlaps_FixedTime}
\end{figure}

To get a deeper understanding of the nature of these two states, as well as why the ground state appears to be clearly resolved without the charmonium operators while the first excitation needs them, we can use the GEVP vectors to access the overlaps between the states created by our different operators and the energy eigenstates \cite{Dudek-2010, Dudek-2008}. Fig. \ref{fig:Overlaps_FixedTime} shows these overlaps at fixed time $t=2a$, for ground state (first column in each pair) and first excitation (second column in each pair). We take the absolute value of the overlaps and normalize them such that the sum of the overlaps per operator is equal to $1$. These overlaps are expected to be time-independent up to exponentially suppressed contributions \cite{Blossier} and we observe this when comparing them across different values of time before the error becomes too large. 
With these results we can make some qualitative statements about the composition of our energy eigenstates. First, the glueball operators with $i=1,2,3,4$ contribute significantly more to the ground state than to the first excitation. The fifth one seems equally distributed for both states. Since the first three glueball operators couple strongly to $J=0$ and this is the lightest glueball, this preference for the ground state is expected. The fourth one coupled strongly to the first three and therefore likely has a sizable $J=0$ contribution, which is consistent with the large overlap onto the ground state. The fifth one does not have a particular preference towards the two states we resolve so it probably has a larger overlap onto a higher state we do not consider here, which would be consistent with a dominant $J=4$ content. For the charmonium operators the pattern of preference is not as marked as for the glueball operators, however it is clear there are at least two which contribute significantly more to the first excitation than to the ground state while the remaining three are approximately equally distributed. The contributions from the former two can help explain why we need charmonium operators to resolve this first excitation which could not be seen clearly by the glueball operators alone. We cannot compare overlaps between different operators onto a same state due to unknown operator-dependent renormalization factors. Based on these overlaps, we identify the ground state to be a mostly gluonic state while the first excitation is mostly mesonic.

\section{Conclusions and outlook}
In this work we presented a systematic study of improved mesonic and gluonic operators to extract the low-lying scalar iso-vector and iso-scalar spectrum in two-flavour QCD. For mesons, we combined the method of optimal distillation profiles with derivative-based operators, widely-used in single- and multi-hadron studies by the Hadron Spectrum collaboration (HadSpec) and observed several key improvements. First, using only the simplest operator with 7 different profiles in a GEVP yielded ground state effective masses which converged much faster to a plateau than the ones coming from a GEVP involving the different derivative-based operators with standard distillation, i.e. constant profile for each one. While including all the available derivative-based operators resulted in even faster convergence, the sole inclusion of profiles had the largest effect. Since carefully chosen derivative-based operators can grant access to states dominated by higher values of $J$ from the continuum, combining suitably chosen derivative-based operators with distillation profiles is the most systematically complete approach to resolve the spectrum. A careful choice of operators for each symmetry channel is fundamental even before introducing the distillation profiles; in the iso-scalar case using standard distillation the one-derivative operator had considerably less excited-state contamination than the zero-derivative one, completely different to the iso-vector case where the effective masses behaved very similarly. This result emphasizes how what is optimal for one channel might be far from being so for another channel.
\par
On the gluonic side, we presented a novel implementation of glueball-like operators built to resemble combinations of the chromo-magnetic field and its gauge-covariant derivatives which have not been systematically used and studied since they were first presented in \cite{Chen-2005, Liu-2000, Liu-2001, Liu-2002}. This construction allowed us to build continuum-like operators with several advantages. Their explicit expressions in term of the chromo-magnetic component provide a more intuitive approach to model glueball-like states on the lattice. By choosing linearly independent operators at different mass-dimensions we obtain a better conditioned basis than spatial Wilson loops. Saturating each mass-dimension is a more systematic way to build independent operators than spatial Wilson loops. In particular, thanks to the non-trivial spatial structure generated by the derivatives, different operators remain independent from each other even after applying 3D link smearing. This directly contrasts the near-degeneracies exhibited by different Wilson loop shapes under the effects of smearing, which was an issue for previous lattice studies \cite{Sakai-2023, Barca2024}. This non-trivial structure leads to a significant spatial size. The use of spatial derivatives in this work made our operators $4a \times 4a$ in size even before including the effects of smearing. As with the meson operators of HadSpec, an increasing number of derivatives, as well as chromo-magnetic components, can be included in the operators by use of well-known Clebsch-Gordan coefficients of either $SO(3)$ or the cubic group, depending on our choice of basis. While we chose the cubic group approach, we still saw different operators were dominated by different $J$ contributions. This resulted in a correlation matrix with a nearly block-diagonal structure, analogous to those obtained by HadSpec using their meson operators built via subduction from $SO(3)$ to the cubic group \cite{Dudek-2010}. Because of this feature, the HadSpec strategy for assigning values of $J$ to the spectrum can straightforwardly be applied using the glueball-like operators implemented in this work. Doing so, we identified the lightest iso-scalar state in our calculation as $J^{PC}=0^{++}$ consisting predominantly of gluons. While we showed that a basis of spatial Wilson loops can recover the same spectrum as these continuum-like operators, their linear independence and strong coupling to definite $J$ make them a better choice of basis. These alternative operators are appealing not only due to the mentioned convenient physical properties, but also thanks to their ease of implementation. By using a clover-definition of the chromo-magnetic field and a nearest-neighbor one for the derivatives instead of linear combinations of Wilson loops, parallelized implementations are straight-forward and are specially beneficial for large enough lattices. Complex spatial structure is achieved without needing to hard-code arbitrarily complicated Wilson loop shapes, whose large extent might also make parallelization difficult or even unfeasible. Since the operators are built with already fixed lattice angular momentum, parity and charge conjugation quantum numbers, the additional work of projecting them onto these fixed quantum numbers which is necessary for spatial Wilson loops is completely avoided. These features, particularly the simplicity in construction even for the case of non-zero spatial momentum, can make these operators useful for recent studies of glueball such as \cite{Abbott-2026}. 
\par
By combining these improved mesonic and gluonic operators we could resolve the two lowest iso-scalar states in our setup, which neither type of operator could do as well by itself, as well as quantify the contributions of each type of operator to these energy eigenstates. Since we observed a larger contribution of the glueball-like operators, particularly those coupling strongly to $J=0$, to the ground state we conclude this is a mostly gluonic one. This was supported also by the the meson operators having larger overlaps with the first excitation rather than this ground state. Based on the preference of the meson operators for the first excitation, as well as the very little contribution it receives from the $J=0$ gluonic operators, we conclude that this excitation is mostly mesonic. While our nonphysical quark masses do not allow a comparison with the meson spectrum in nature, at our relatively fine lattice spacing we find this mostly-gluonic iso-scalar ground state close to the pure-gauge predictions of $1.8$ GeV \cite{Peardon-1999}.
\par
The improvements presented in this work can be applied well beyond what was shown here. HadSpec currently uses an extensive basis of operators for scattering studies where reliable finite-volume spectrum determinations are fundamental \cite{Cheung-2020, Hansen-2021, Gayer-2021, Wilson-2024, Wilson-2024A, Yeo-2024, Whyte-2025, Pitanga-2025}. While derivative-based operators are definitely necessary, some of them might be close to redundant and therefore could be replaced by extending a minimal set with distillation profiles. Alternatively, one could determine the optimal profiles for each single-hadron operator and then introduce them in the corresponding component of the multi-particle operators. Other studies which rely on shorter lists of operators, e.g. only $\gamma_5$, $\gamma_i$ and other combinations of Dirac matrices, could very cheaply extend their basis with distillation profiles as well \cite{Li-2025,Li-2025A, Shrimal-2025, Prelovsek-2025, Meng-2025, Sadl-2025}. Glueball studies, whether in pure-gauge or with dynamical quarks, can benefit from the features of the glueball-like operators explored here; the rather arbitrary choices of Wilson loops can be replaced by the systematic combinations of chromo-magnetic components and their derivatives, allowing strong coupling to states of interest and spin-identification which is already widely used for mesons, as well as simplifying the human effort behind building and implementing the operators. Suitable operator construction is a fundamental issue in lattice hadron spectroscopy, since unreliably plateaus, or missing states in the worst case, can severely hinder calculations even when there is good statistical precision. With the methods explored here we aimed to extend the tools available for reliable spectrum determinations via optimized operators and will use them in future studies of the scalar glueball, where meson and glueball operators must be mixed at different lattice volumes as a first step towards a scattering analysis.

\begin{acknowledgments}
J.A.U.-N. kindly thanks Nikolai Husung for all his help using the \textit{opbasis} Python package~\cite{Husung-2025}. The authors gratefully acknowledge the Gauss Centre for Supercomputing e.V. (www.gauss-centre.eu) for funding this project by providing computing time on the GCS Supercomputer SuperMUC-NG at Leibniz Supercomputing Centre (www.lrz.de). We also thank Juelich Supercomputing Centre (JSC) for allocating space on the storage system JUST (project ID hwu35).
J.A.U.-N. acknowledges support from a Research Ireland (Science Foundation Ireland) Frontiers for the Future Project award [grant number SFI-21/FFP-P/10186]. M.P. was supported by the European Union’s Horizon 2020 research and innovation programme under grant agreement 824093 (STRONG-2020). The work is supported by the German Research Foundation (DFG) research unit FOR5269 "Future methods for studying confined gluons in QCD". The sole responsibility for the content of this publication lies with the authors. 
\end{acknowledgments}
\clearpage

\bibliography{apssamp}



\end{document}